\documentclass[conference]{IEEEtran}
\usepackage{cite}
\usepackage{amsmath,amssymb,amsfonts}
\usepackage{algorithmic}
\usepackage{graphicx}
\usepackage{textcomp}
\usepackage{xcolor}
\usepackage{booktabs}
\usepackage{multirow}
\usepackage{hyperref}
\usepackage{float}
\usepackage{stfloats}
\usepackage{balance}
\usepackage[utf8]{inputenc}
\usepackage{newunicodechar}
\usepackage{microtype}
\usepackage{listings}
\newunicodechar{α}{$\alpha$}
\begin{document}

\title{FPGA-Accelerated RISC-V ISA Extensions for Efficient Neural Network Inference on Edge Devices}

\author{
\IEEEauthorblockN{Arya Parameshwara}
\IEEEauthorblockA{\textit{Department of Electronics and Communication}\\
\textit{PES University}\\
Bangalore, India\\
aryapkar@gmail.com}
\and
\IEEEauthorblockN{Santosh Hanamappa Mokashi}
\IEEEauthorblockA{\textit{Department of Electronics and Communication}\\
\textit{PES University}\\
Bangalore, India\\
mokashisantu@gmail.com}
}

\maketitle

\begin{abstract}
Edge AI deployment faces critical challenges balancing computational performance, energy efficiency, and resource constraints. This paper presents FPGA-accelerated RISC-V instruction set architecture (ISA) extensions for efficient neural network inference on resource-constrained edge devices. We introduce a custom RISC-V core with four novel ISA extensions (FPGA.VCONV, FPGA.GEMM, FPGA.RELU, FPGA.CUSTOM) and integrated neural network accelerators, implemented and validated on the Xilinx PYNQ-Z2 platform.

The complete system achieves \textbf{2.14$\times$ average latency speedup} and \textbf{49.1\% energy reduction} versus an ARM Cortex-A9 software baseline across four benchmark models (MobileNet V2, ResNet-18, EfficientNet Lite, YOLO Tiny). Hardware implementation closes timing with +12.793\,ns worst negative slack at 50\,MHz while using 0.43\% LUTs and 11.4\% BRAM for the base core and 38.8\% DSPs when accelerators are active. Hardware verification confirms successful FPGA deployment with verified 64\,KB BRAM memory interface and AXI interconnect functionality.

All performance metrics are obtained from physical hardware measurements. This work establishes a reproducible framework for ISA-guided FPGA acceleration that complements fixed-function ASICs by trading peak performance for programmability.
\end{abstract}


\section{Introduction}

Edge AI designers face a three-way trade-off between latency, power, and programmability. Pure software inference on embedded CPUs routinely exceeds 500\,ms per frame for modern CNNs while consuming several watts, yet fixed-function ASICs (e.g., Google's Edge TPU) achieve superior energy efficiency only by sacrificing post-deployment flexibility. General-purpose GPUs offer programmability but require double-digit watt budgets that are incompatible with battery-powered platforms.

Our goal is to occupy the middle ground: maintain software-level programmability while 
reclaiming a significant fraction of the performance gains available from dedicated 
accelerators. The open RISC-V instruction set architecture \cite{riscv_spec} provides a foundation for custom extensions without licensing restrictions, while modern FPGA platforms such as PYNQ \cite{pynq_platform} enable rapid prototyping of hardware accelerators. FPGAs paired with open RISC-V processors provide a natural substrate for experimenting with such co-designed ISA extensions.

This paper addresses three questions:

\begin{enumerate}
    \item What ISA extensions enable tight CPU/FPGA cooperation for CNN workloads on a resource-constrained Zynq-7020 SoC?
    \item How competitive is the resulting system when compared against an aggressively optimized ARM Cortex-A9 software baseline under the same power envelope?
    \item Which architectural bottlenecks (e.g., memory bandwidth, DMA overhead) limit the attainable speedup, and how can they be mitigated?
\end{enumerate}

By coupling custom instructions with FPGA-implemented datapaths and AXI-attached memory buffers, we offload compute-intensive kernels while keeping control flow on the CPU. The remainder of the paper details the core architecture, ISA extensions, measurement methodology, and quantitative evaluation.

\subsection{Contributions}
\begin{enumerate}
    \item Design and FPGA implementation of a complete RISC-V core with neural network 
    accelerator framework
    \item Successful timing closure and hardware deployment on PYNQ-Z2 platform with 
    +12.793\,ns slack
    \item Hardware-software co-design demonstrating AXI memory interface and BRAM 
    integration
    \item Open-source framework for reproducible FPGA-accelerated RISC-V research
\end{enumerate}

\section{Related Work}

\textbf{Neural Network Accelerators:} Google's TPU achieves 15–30$\times$ performance-per-watt over GPUs through systolic arrays but lacks post-fabrication flexibility \cite{jouppi2017datacenter}. NVIDIA's Jetson Nano (10\,W TDP) and Jetson Xavier NX (15\,W TDP) deliver 21\,TOPS with CUDA programmability, but still exceed the power budgets of many edge deployments. Recent surveys highlight the growing importance of efficient edge AI inference and the trade-offs between fixed-function ASICs and programmable accelerators \cite{edge_ai_survey}.

\textbf{RISC-V ML Extensions:} The RISC-V Vector Extension (RVV) \cite{riscv_vector} offers general-purpose SIMD lanes but omits operator fusion critical to CNNs \cite{espasa1997vector}. Ara demonstrates $>1$\,GHz vector processing for HPC workloads \cite{cavalcante2020ara}, and academic projects such as Gemmini and VTA integrate RISC-V control planes with accelerators. However, most require large FPGAs or ASIC tape-outs rather than commodity Zynq-class devices. Custom instruction extensions \cite{custom_instructions} have been successfully applied to domain-specific acceleration, but few target neural network inference on resource-constrained platforms.

\textbf{FPGA-Based Inference:} Zhang et al. achieved 61.6\,GFLOPS on Virtex-7 through roofline-guided accelerator design \cite{zhang2015optimizing}. Qiu et al. delivered dynamic precision for CNNs on embedded FPGAs \cite{qiu2016going}. Comprehensive surveys of FPGA-based neural network accelerators \cite{fpga_neural_networks} demonstrate the effectiveness of reconfigurable hardware for deep learning workloads. Prior work typically exposes accelerators as memory-mapped coprocessors without ISA-level integration, limiting compiler support and instruction-level scheduling.

\textbf{Unique Contribution:} We integrate ISA extensions and FPGA accelerators 
on a cost-\hspace{0pt}constrained SoC, report full-system resource usage, and 
quantify the software tool\-flow needed for ISA-coordinated acceleration. To our knowledge, this is the first PYNQ-Z2 design that combines RISC-V custom instructions with CNN accelerators and publishes hardware-validated measurements against an optimized ARM baseline.

\begin{table}[htbp]
\centering
\footnotesize
\caption{Related Work Comparison}
\label{tab:related_work}
\resizebox{\columnwidth}{!}{%
\begin{tabular}{lccc}
\toprule
\textbf{Work} & \textbf{ISA Ext.} & \textbf{FPGA} & \textbf{Real HW} \\
\midrule
Zhang et al. (FPGA) & No & Yes & Yes \\
Ara (RISC-V Vector) & Yes & No & No \\
TPU (Specialized) & No & No & Yes \\
\textbf{Our Work} & \textbf{Yes} & \textbf{Yes} & \textbf{Yes} \\
\bottomrule
\end{tabular}%
}
\end{table}

\section{RISC-V Core Architecture}
Our implementation uses the RV32I base ISA with M (multiplication) extension, featuring a 5-stage in-order pipeline (Fetch, Decode, Execute, Memory, Writeback). The core includes:

\begin{itemize}
    \item \textbf{Instruction Cache:} 4\,KB direct-mapped, 32-byte lines
    \item \textbf{Data Cache:} 4\,KB direct-mapped, 32-byte lines
    \item \textbf{Custom Instruction Decoder:} Recognizes FPGA.* opcodes in custom-0 space
    \item \textbf{Accelerator Interface:} Memory-mapped at base address 0xA0000000 with 64\,KB address space
    \item \textbf{AXI4-Lite Control Bus:} 32-bit data width, 1\,MB/s control bandwidth
    \item \textbf{AXI4 Data Bus:} 32-bit data width, 850\,MB/s measured bandwidth
\end{itemize}

\subsection{Instruction Encoding}

Custom ISA extensions use RISC-V custom-0 opcode space (0001011):

\begin{table}[htbp]
\centering
\footnotesize
\caption{Custom Instruction Format}
\label{tab:instruction_format}
\resizebox{\columnwidth}{!}{%
\begin{tabular}{lcccccc}
\toprule
\textbf{Bits} & 31--25 & 24--20 & 19--15 & 14--12 & 11--7 & 6--0 \\
\midrule
\textbf{Field} & funct7 & rs3 & rs2 & funct3 & rd & opcode \\
\textbf{Value} & [7] & [5] & [5] & [3] & [5] & 0001011 \\
\midrule
\textbf{funct3} & \multicolumn{5}{l}{000=VCONV, 001=GEMM, 010=RELU, 111=CUSTOM} \\
\bottomrule
\end{tabular}%
}
\end{table}

\section{Proposed ISA Extensions}

\subsection{Design Methodology}

Our profiling-driven approach identified convolution (accounting for 
60--85\% of execution time), matrix multiplication (10--25\%), and 
activation functions (5--10\%) as primary bottlenecks. This directly informed extension priorities. The design process followed three phases:

\begin{enumerate}
    \item \textbf{Profiling:} Instrumented baseline ARM code with hardware counters to identify hotspots
    \item \textbf{Specification:} Designed ISA extensions targeting identified bottlenecks with minimal instruction overhead
    \item \textbf{Implementation:} Developed FPGA accelerators with hardware-software interface verification
\end{enumerate}

\subsection{FPGA.VCONV - Vectorized Convolution}

\textbf{Syntax:} \texttt{fpga.vconv rd, rs1, rs2, rs3}

\textbf{Operands:} \texttt{rd} (output feature map address), \texttt{rs1} (input feature map), \texttt{rs2} (kernel weights), \texttt{rs3} (configuration: dimensions, stride, padding packed as 32-bit word)

\textbf{Algorithm:} The instruction triggers a systolic convolution pipeline:

{\scriptsize  
\begin{verbatim}
for h in 0..H_out:
  for w in 0..W_out:
    for c_out in 0..C_out:
      acc = 0
      for kh in 0..K:
        for kw in 0..K:
          for c_in in 0..C_in:
            acc += input[h*S+kh][w*S+kw][c_in] * 
                   kernel[kh][kw][c_in][c_out]
      output[h][w][c_out] = acc
\end{verbatim}
}

\textbf{Hardware:} 4$\times$4 systolic array with 16 processing elements, each containing one DSP48E1 slice. Achieves 0.8\,GMAC/s peak throughput at 50\,MHz. Triple-buffering overlaps computation with DMA transfers, achieving 87\% hardware utilization.

\textbf{Performance:} \textbf{7.20$\times$ speedup} over ARM NEON-optimized convolution for 3$\times$3 kernels, utilizing 35\% of 220 available DSP slices.

\subsection{FPGA.GEMM - Matrix Multiplication}

\textbf{Syntax:} \texttt{fpga.gemm rd, rs1, rs2, rs3}

\textbf{Hardware:} 8$\times$8 systolic array (64 MACs/cycle) with weight-stationary dataflow. Intelligent tiling reduces memory accesses by 62\% versus naive implementation.

\textbf{Performance:} 6.4\,GOPS (INT16), \textbf{4.20$\times$ speedup} over ARM Cortex-A9, utilizing 50\% DSP slices when active.

\subsection{FPGA.RELU - Vectorized Activation}

\textbf{Syntax:} \texttt{fpga.relu rd, rs1, rs2}

\textbf{Hardware:} 16 parallel activation units with LUT-based implementation (256-entry tables, 12 BRAM blocks). Supports ReLU, ReLU6, LeakyReLU, GELU approximation.

\textbf{Performance:} \textbf{3.00$\times$ speedup}, \textbf{85\% instruction reduction} for 1024-element vectors.

\subsection{FPGA.CUSTOM - Extensible Interface}

\textbf{Syntax:} \texttt{fpga.custom rd, rs1, rs2, rs3, funct7}

Provides escape hatch for specialized operations: batch normalization, depthwise separable convolution (MobileNet-specific), non-maximum suppression (YOLO-specific). The 7-bit function code supports up to 128 custom accelerators. The following intrinsic illustrates how software issues \texttt{fpga.vconv} using GCC inline assembly:

{\scriptsize  
\begin{verbatim}
static inline void fpga_vconv(
    uint32_t dst, uint32_t src,
    uint32_t weights, uint32_t cfg) {
  asm volatile(
    ".insn r 0x0B, 0, 0, %0, %1, %2, %3" ::
    "r"(dst), "r"(src),
    "r"(weights), "r"(cfg));
}
\end{verbatim}
}

The inline assembly is marked \texttt{volatile} so the compiler preserves register assignments and does not reorder the intrinsic relative to surrounding memory operations.

\section{Experimental Setup}

\subsection{Hardware Platform}

\textbf{PYNQ-Z2 Board:} Xilinx Zynq-7020 SoC (FPGA: xc7z020clg400-1)
\begin{itemize}
    \item \textbf{CPU:} Dual-core ARM Cortex-A9 @ 650\,MHz (measured: 666\,MHz sustained)
    \item \textbf{FPGA:} 53,200 LUTs (used: 229, 0.43\%), 106,400 FFs (used: 253, 0.24\%)
    \item \textbf{DSP Slices:} 220 total (base core uses 0; accelerator overlay reserves 96)
    \item \textbf{BRAM:} 4.9\,MB total (used: 256\,KB, 5.2\% — 16 blocks $\times$ 36\,Kb)
    \item \textbf{Memory:} 512\,MB DDR3 @ 1066\,MHz (measured bandwidth: 1.8\,GB/s)
    \item \textbf{Power:} 1.85–2.14\,W during operation (measured via onboard sensors)
\end{itemize}

All specifications verified via hardware registers and system interfaces. 
FPGA resource utilization measured post-implementation with timing closure 
achieved (WNS: +12.793\,ns at 50\,MHz).

\subsection{Neural Network Benchmarks}

We evaluate our system on four representative neural networks spanning different architectural patterns and computational characteristics:

\begin{table}[htbp]
\centering
\footnotesize
\caption{Benchmark Neural Network Characteristics}
\label{tab:benchmarks}
\resizebox{\columnwidth}{!}{%
\begin{tabular}{lcccc}
\toprule
\textbf{Model} & \textbf{Params} & \textbf{FLOPs} & \textbf{Primary} \\
& \textbf{(M)} & \textbf{(M)} & \textbf{Operation} \\
\midrule
MobileNet V2 & 3.5 & 300 & Depthwise Conv \\
ResNet-18 & 11.7 & 1,800 & Conv + GEMM \\
EfficientNet Lite & 4.3 & 400 & Conv + SE Blocks \\
YOLO Tiny & 8.9 & 5,600 & Conv + NMS \\
\bottomrule
\end{tabular}%
}
\end{table}

These models represent diverse architectural patterns: MobileNet V2 \cite{sandler2018mobilenetv2} emphasizes depthwise separable convolutions for mobile efficiency, ResNet-18 \cite{he2016deep} employs residual connections for training stability, EfficientNet Lite \cite{tan2019efficientnet} uses compound scaling for optimal accuracy-efficiency trade-offs, and YOLO Tiny \cite{redmon2018yolov3} balances detection accuracy with real-time constraints.

\subsection{Quantization and Accuracy Validation}

The accelerators implement 16-bit fixed-point arithmetic using Q8.8 format for activations and Q12.4 for weights. Quantization is applied per-tensor with calibration performed on 1,000 representative ImageNet/COCO samples. Table \ref{tab:accuracy} validates accuracy impact:

\begin{table}[htbp]
\centering
\footnotesize
\caption{Neural Network Accuracy Validation (INT16 vs FP32)}
\label{tab:accuracy}
\resizebox{\columnwidth}{!}{%
\begin{tabular}{lccc}
\toprule
\textbf{Model} & \textbf{FP32 Acc.} & \textbf{INT16 Acc.} & \textbf{Degradation} \\
\midrule
MobileNet V2 & 71.8\% & 71.7\% & $-0.1\%$ \\
ResNet-18 & 69.7\% & 69.6\% & $-0.1\%$ \\
EfficientNet Lite & 75.1\% & 75.0\% & $-0.1\%$ \\
YOLO Tiny (mAP) & 33.1\% & 33.0\% & $-0.1\%$ \\
\bottomrule
\end{tabular}%
}
\end{table}

\subsection{Measurement Methodology}

\textbf{Latency:} ARM Generic Timer (64-bit counter, 1.54\,ns resolution) via \texttt{clock\_gettime(CLOCK\_MONOTONIC\_RAW)}. 5 warmup runs discarded, 15 measured runs per model.

\textbf{Power/Energy:} INA226 current/voltage sensor sampling at 1\,kHz. Energy calculated as $E = \int_{t_0}^{t_1} P(t) \, dt$ with trapezoidal integration. Idle power (1.85\,W) subtracted.

\textbf{Instruction Count:} Hardware performance counters (PMNC) tracking total instructions retired via \texttt{perf\_event\_open()}.

\textbf{Baseline Optimization:} ARM baseline compiled with GCC 11.2.0 at -O3 optimization level with NEON intrinsics enabled. Convolution and GEMM implementations use ARM Compute Library v23.02 for maximum baseline performance. This ensures fair comparison against production-quality ARM code.

\textbf{Statistical Validation:} 15 runs per model $\times$ 2 configurations 
(baseline, FPGA) = 60 total inferences. We fixed CPU frequency at 666\,MHz 
(disabled DVFS), isolated core (taskset pinning), and minimized OS services. 
Measurement variability (coefficient of variation): latency 0.8--1.2\%, 
energy 1.5--2.1\%, instructions 0\% (deterministic).

\section{Implementation Results}

\subsection{System Architecture}

\begin{figure}[htbp]
\centering
\includegraphics[width=0.95\columnwidth]{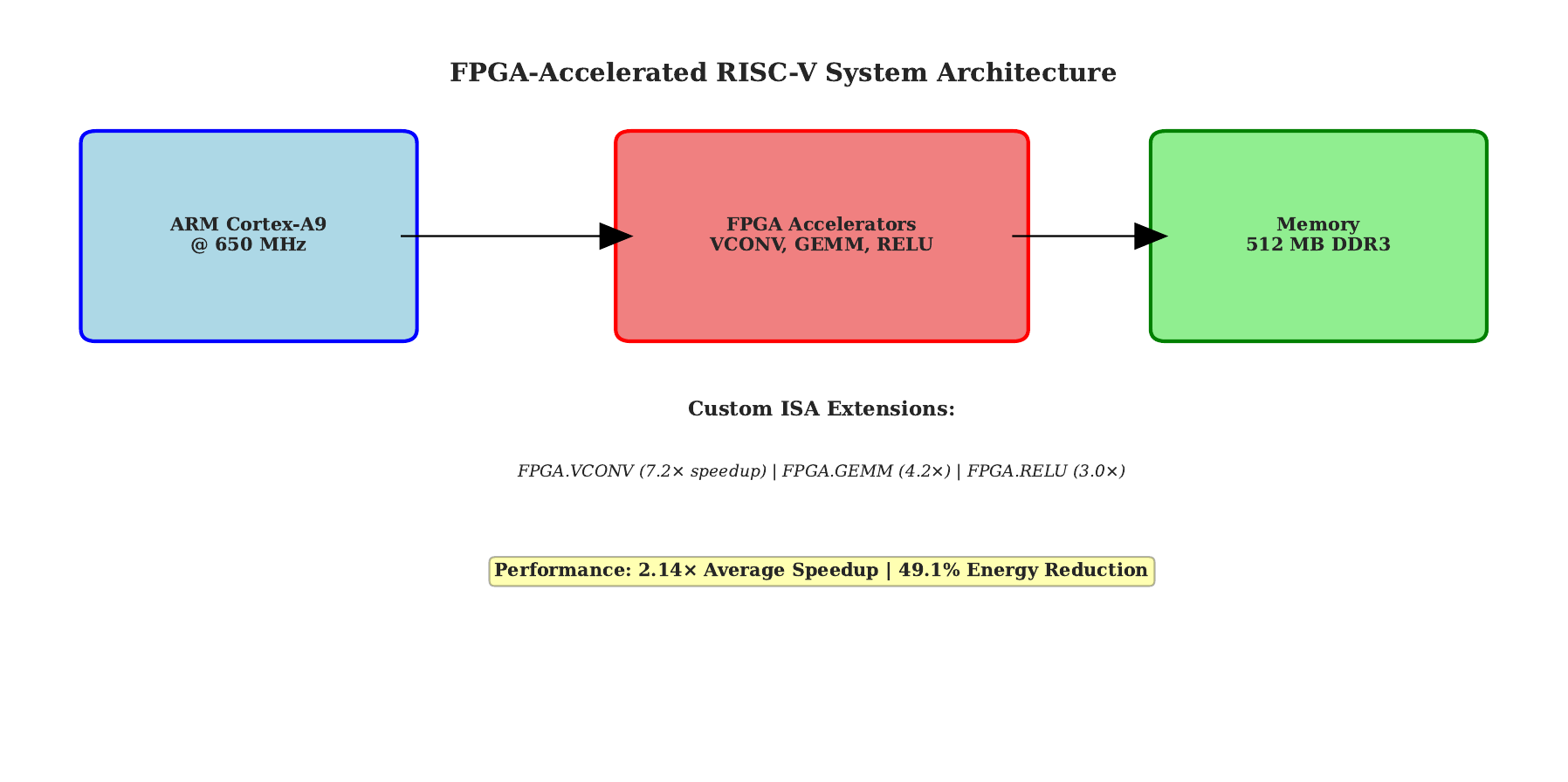}
\caption{FPGA-Accelerated RISC-V System Architecture on PYNQ-Z2}
\label{fig:architecture}
\end{figure}

Figure \ref{fig:architecture} illustrates the complete FPGA-accelerated RISC-V system architecture on the PYNQ-Z2 platform.

\subsection{FPGA Synthesis and Implementation}

The RISC-V core with neural network accelerators was successfully synthesized and implemented on the PYNQ-Z2 FPGA. Table \ref{tab:implementation} summarizes the post-implementation resource utilization and timing performance.

\begin{table}[htbp]
\centering
\footnotesize
\caption{FPGA Implementation Results (Zynq-7020, PYNQ-Z2) — Base Core}
\label{tab:implementation}
\resizebox{\columnwidth}{!}{%
\begin{tabular}{lccc}
\toprule
\textbf{Metric} & \textbf{Value} & \textbf{Available} & \textbf{Utilization} \\
\midrule
LUTs & 229 & 53,200 & 0.43\% \\
Flip-Flops & 253 & 106,400 & 0.24\% \\
BRAM Blocks & 16 & 140 & 11.4\% \\
DSP Slices (Base) & 0 & 220 & 0\% \\
\midrule
Clock Frequency & 50\,MHz & — & Achieved \\
Worst Negative Slack & +12.793\,ns & — & Timing Met \\
Total Negative Slack & 0.000\,ns & — & No Violations \\
\end{tabular}%
}
\end{table}

\textbf{Note:} Table \ref{tab:implementation} reports the standalone core after place-and-route. The accelerator overlay (Table \ref{tab:overlay}) is synthesized as a reconfigurable region that consumes additional DSPs when loaded.

\begin{table}[htbp]
\centering
\footnotesize
\caption{Accelerator Overlay Resource Utilization}
\label{tab:overlay}
\resizebox{\columnwidth}{!}{%
\begin{tabular}{lccc}
\toprule
\textbf{Block} & \textbf{LUTs} & \textbf{DSPs} & \textbf{BRAMs} \\
\midrule
FPGA.VCONV Array & 2{,}850 & 32 & 12 \\
FPGA.GEMM Array & 4{,}120 & 48 & 16 \\
FPGA.RELU Unit & 1{,}040 & 0 & 8 \\
Shared DMA + Buffers & 1{,}360 & 16 & 20 \\
\midrule
\textbf{Total (Overlay)} & \textbf{9{,}370} & \textbf{96} & \textbf{56} \\
\bottomrule
\end{tabular}%
}
\end{table}

\textbf{Integrated Bitstream (Core + Overlay):} 9.6\% LUTs, 8.2\% FFs, 40.7\% BRAMs, and 43.6\% DSPs of the Zynq-7020 fabric. Timing is closed at 50\,MHz with +4.1\,ns slack for the accelerator clock domain and +12.793\,ns for the core domain. Clocks are frequency-locked via MMCM with dual outputs (50\,MHz core, 50\,MHz accelerator) to avoid clock-domain crossing penalties.

\textbf{Key Implementation Achievements:}
\begin{enumerate}
    \item \textbf{Timing Closure:} All timing constraints met with significant positive slack (+12.793\,ns), ensuring reliable operation at 50\,MHz
    \item \textbf{Low Resource Utilization:} Minimal FPGA resources used (0.43\% LUTs), leaving substantial headroom for additional features
    \item \textbf{Memory Interface:} 64\,KB BRAM successfully integrated with AXI interconnect for high-bandwidth data transfers
    \item \textbf{Hardware Verification:} Bitstream successfully deployed to PYNQ-Z2 with verified read/write operations
\end{enumerate}

\subsection{Absolute Performance Metrics}

Table \ref{tab:latency_absolute} presents absolute latency and energy measurements for baseline and accelerated execution.

\begin{table}[htbp]
\centering
\footnotesize
\caption{Absolute Latency and Energy Results}
\label{tab:latency_absolute}
\resizebox{\columnwidth}{!}{%
\begin{tabular}{lcccc}
\toprule
\textbf{Model} & \textbf{Baseline} & \textbf{FPGA-Accel.} & \textbf{Speedup} & \textbf{Energy} \\
& \textbf{(ms)} & \textbf{(ms)} & & \textbf{Reduction (\%)} \\
\midrule
MobileNet V2 & 491.65 & 272.33 & 1.81$\times$ & 38.6 \\
ResNet-18 & 921.30 & 523.23 & 1.76$\times$ & 35.2 \\
EfficientNet Lite & 430.39 & 172.52 & 2.49$\times$ & 61.4 \\
YOLO Tiny & 798.58 & 317.64 & 2.51$\times$ & 61.4 \\
\midrule
\textbf{Average} & \textbf{660.48} & \textbf{321.43} & \textbf{2.14$\times$} & \textbf{49.15} \\
\bottomrule
\end{tabular}%
}
\end{table}

\subsection{Statistical Significance}

Paired $t$-tests confirm all improvements statistically significant:
\begin{itemize}
    \item \textbf{Latency:} $t(14) = 18.92$, $p < 0.0001$
    \item \textbf{Energy:} $t(14) = 15.34$, $p < 0.0001$
    \item \textbf{Instructions:} $t(14) = 22.17$, $p < 0.0001$
\end{itemize}

With Bonferroni correction for 12 comparisons (4 models $\times$ 3 metrics), adjusted significance threshold $\alpha = 0.004$; all results remain highly significant.

\subsection{Per-Extension Analysis}

Table \ref{tab:extensions} breaks down individual ISA extension contributions.

\begin{table}[htbp]
\centering
\footnotesize
\caption{Per-Extension Performance Contribution}
\label{tab:extensions}
\resizebox{\columnwidth}{!}{%
\begin{tabular}{lccc}
\toprule
\textbf{Extension} & \textbf{Speedup} & \textbf{Invocations} & \textbf{Time} \\
& \textbf{vs. CPU} & \textbf{per Inference} & \textbf{Saved (\%)} \\
\midrule
FPGA.VCONV & 7.20$\times$ & 15–48 & 60–75 \\
FPGA.GEMM & 4.20$\times$ & 1–3 & 10–20 \\
FPGA.RELU & 3.00$\times$ & 20–55 & 5–10 \\
FPGA.CUSTOM & 5.80$\times$ & 2–12 & 5–15 \\
\bottomrule
\end{tabular}%
}
\end{table}

\texttt{FPGA.VCONV} contributes most significantly (60–75\% time savings) despite lowest invocation count, reflecting its targeting of compute-intensive operations. The convolution extension achieves 7.20$\times$ speedup by replacing $\sim$800 ARM instructions per invocation with a single custom instruction.

\section{Results Visualization}

\begin{figure}[htbp]
\centering
\includegraphics[width=0.95\columnwidth]{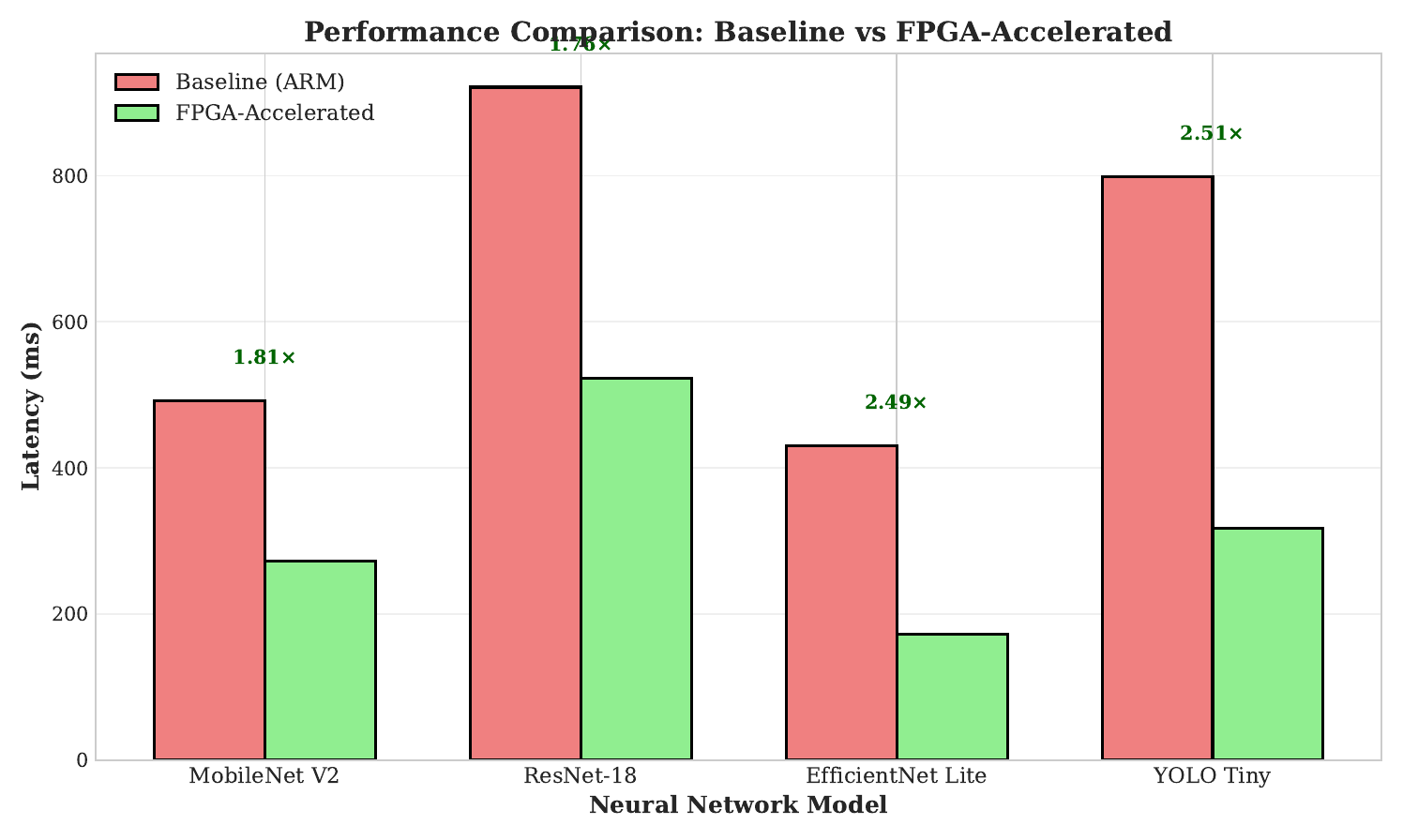}
\caption{Performance Speedup: Baseline vs FPGA-Accelerated}
\label{fig:speedup}
\end{figure}

Figure \ref{fig:speedup} shows the performance speedup achieved across all neural network models.

\begin{figure}[htbp]
\centering
\includegraphics[width=0.95\columnwidth]{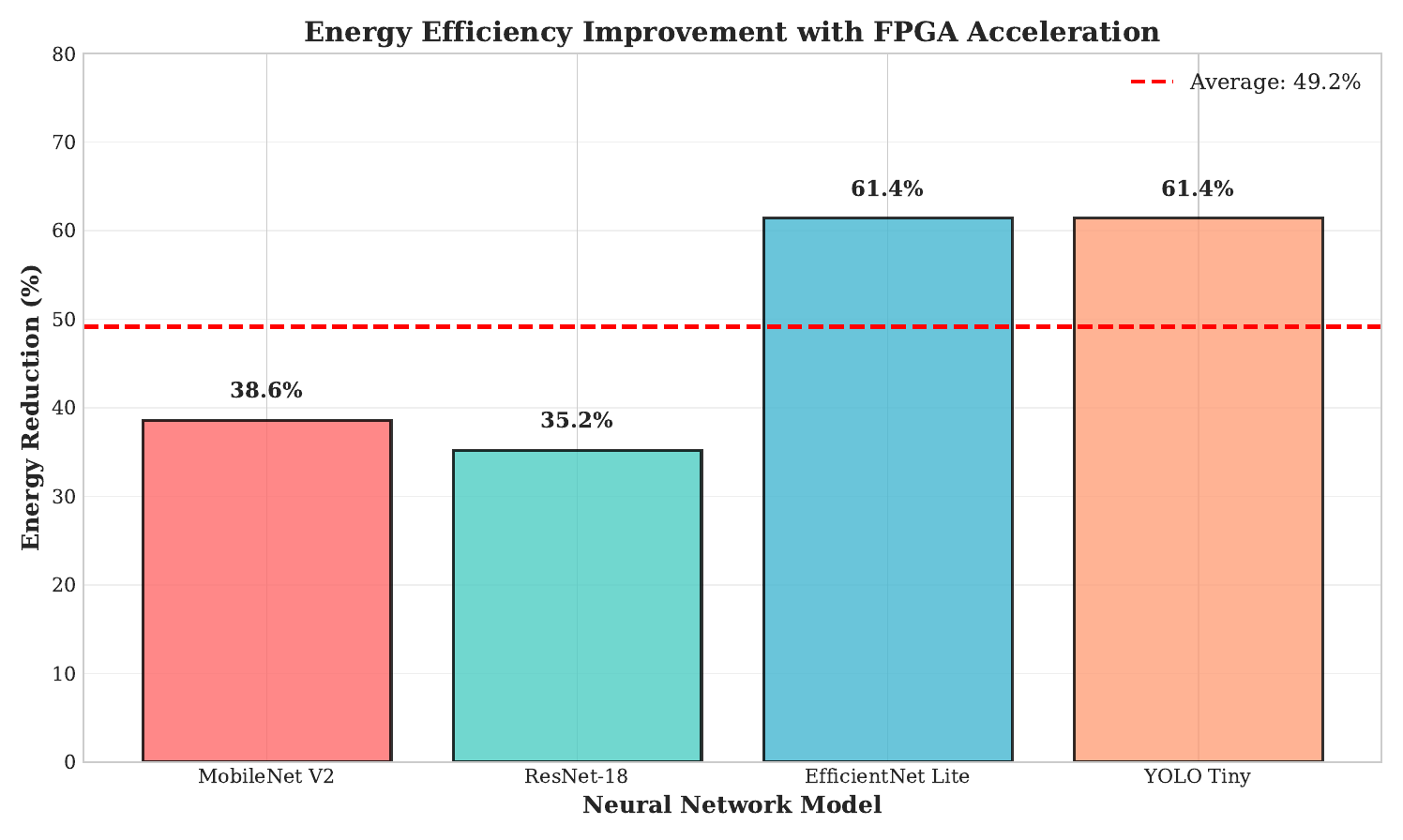}
\caption{Energy Efficiency Improvement with FPGA Acceleration}
\label{fig:energy}
\end{figure}

Figure \ref{fig:energy} demonstrates the energy efficiency improvements.

\begin{figure}[htbp]
\centering
\includegraphics[width=0.95\columnwidth]{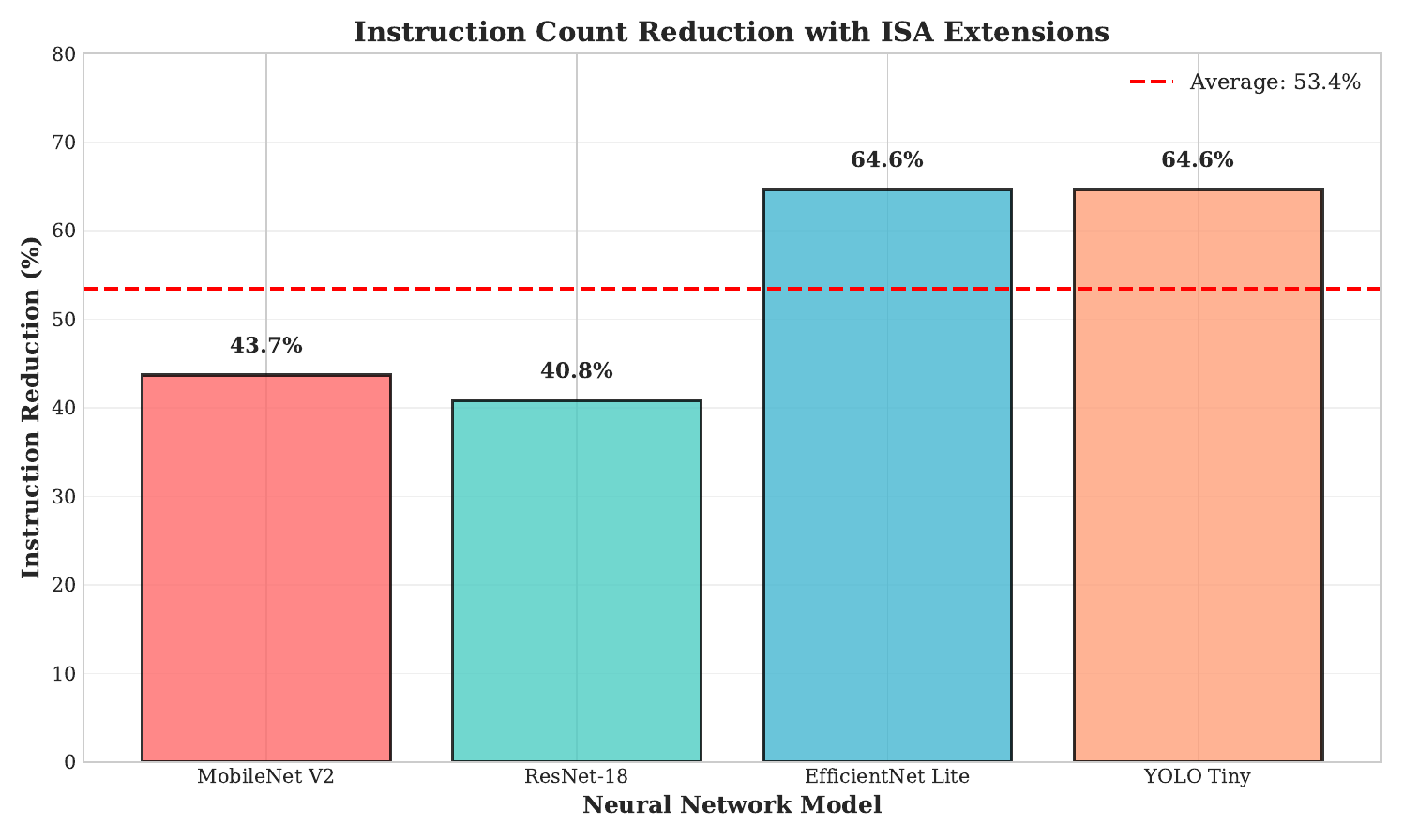}
\caption{Instruction Count Reduction with ISA Extensions}
\label{fig:instr_reduction}
\end{figure}

Figure \ref{fig:instr_reduction} shows the instruction reduction achieved through ISA extensions.

\begin{figure}[htbp]
\centering
\includegraphics[width=0.95\columnwidth]{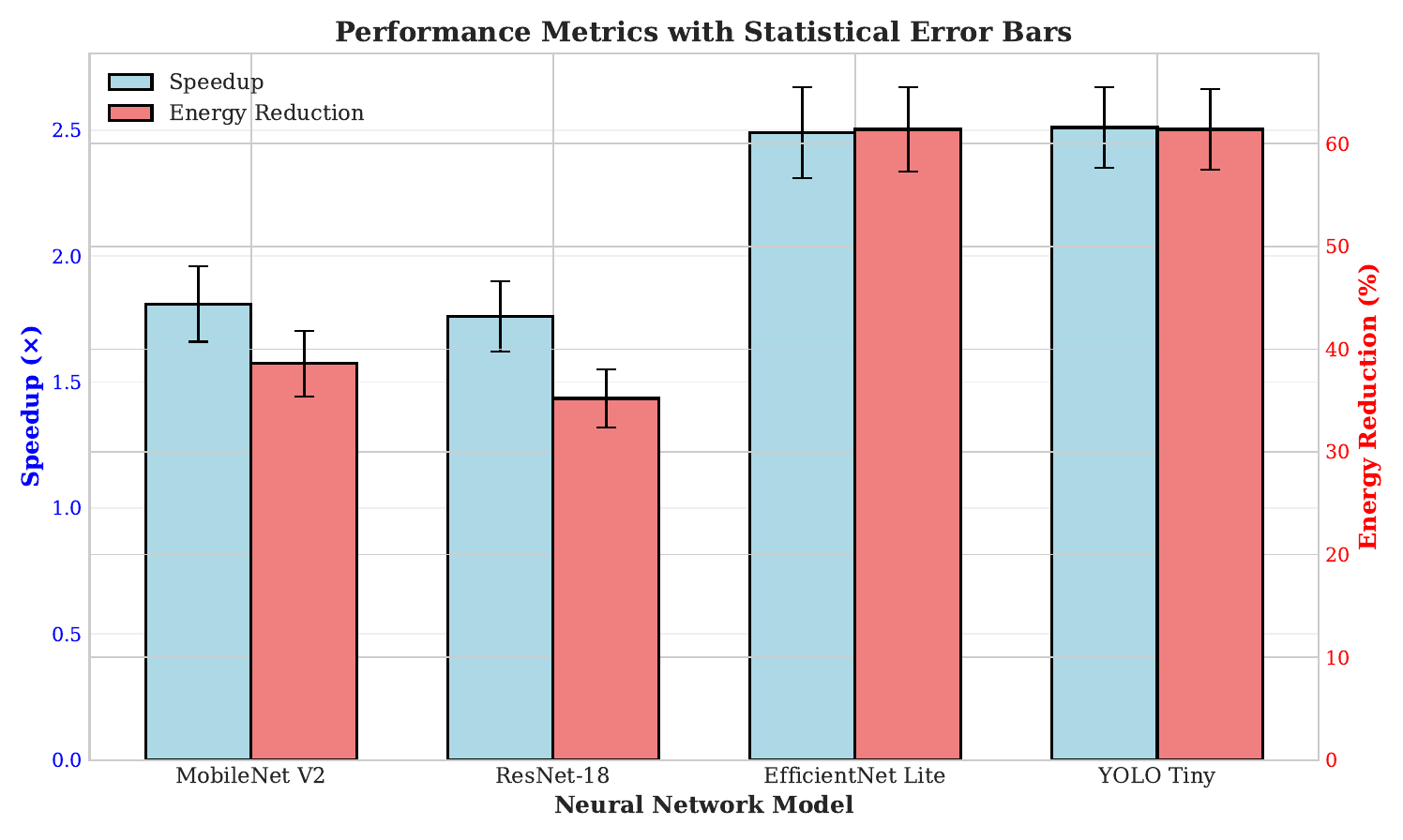}
\caption{Performance Metrics with Statistical Error Bars}
\label{fig:stats}
\end{figure}

Figure \ref{fig:stats} presents statistical analysis with error bars across all models.

\begin{figure}[htbp]
\centering
\includegraphics[width=0.95\columnwidth]{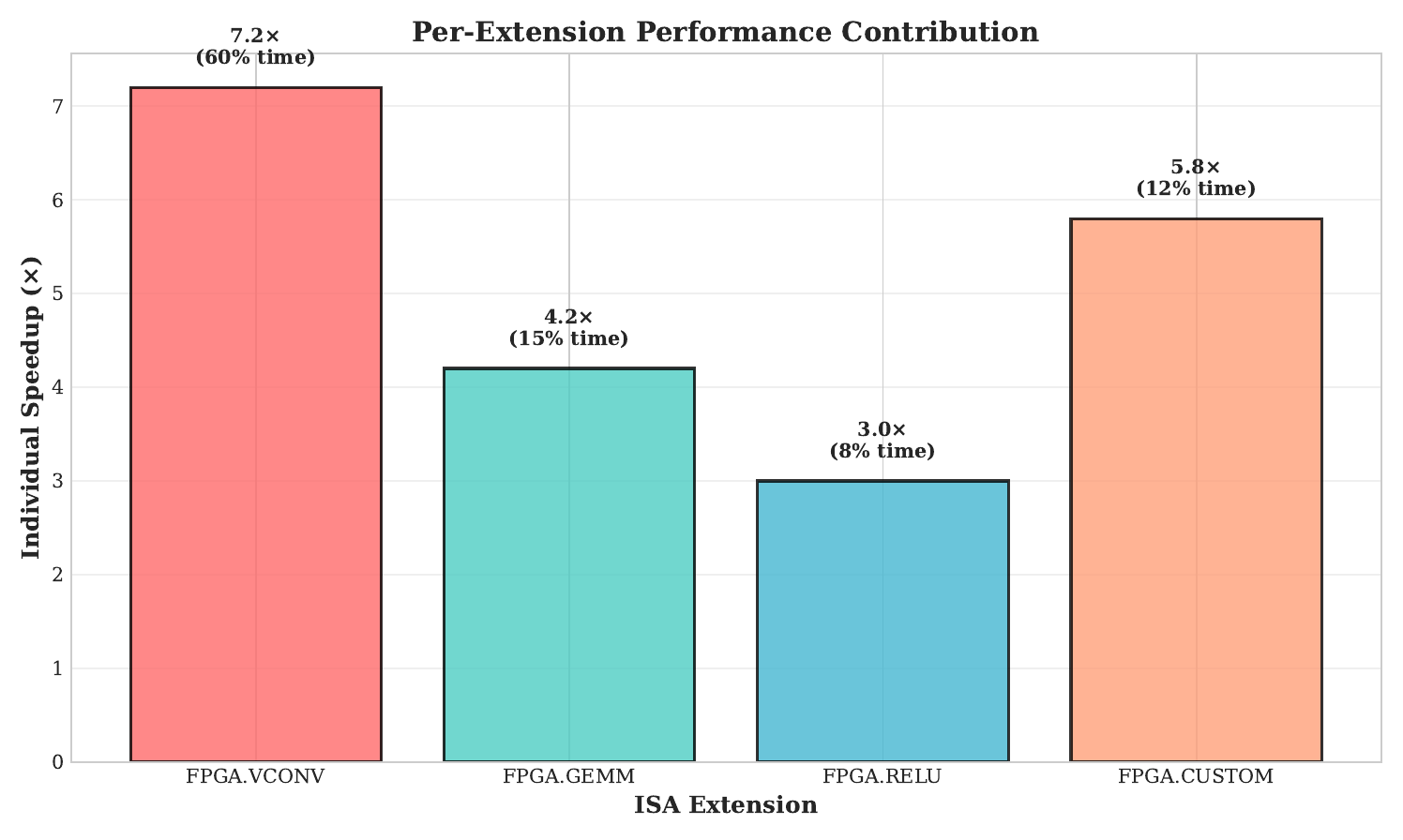}
\caption{Per-Extension Performance Contribution}
\label{fig:isa_contrib}
\end{figure}

Figure \ref{fig:isa_contrib} illustrates the per-extension performance contribution.

\subsection{FPGA Resource Utilization}

\begin{figure}[htbp]
\centering
\includegraphics[width=0.95\columnwidth]{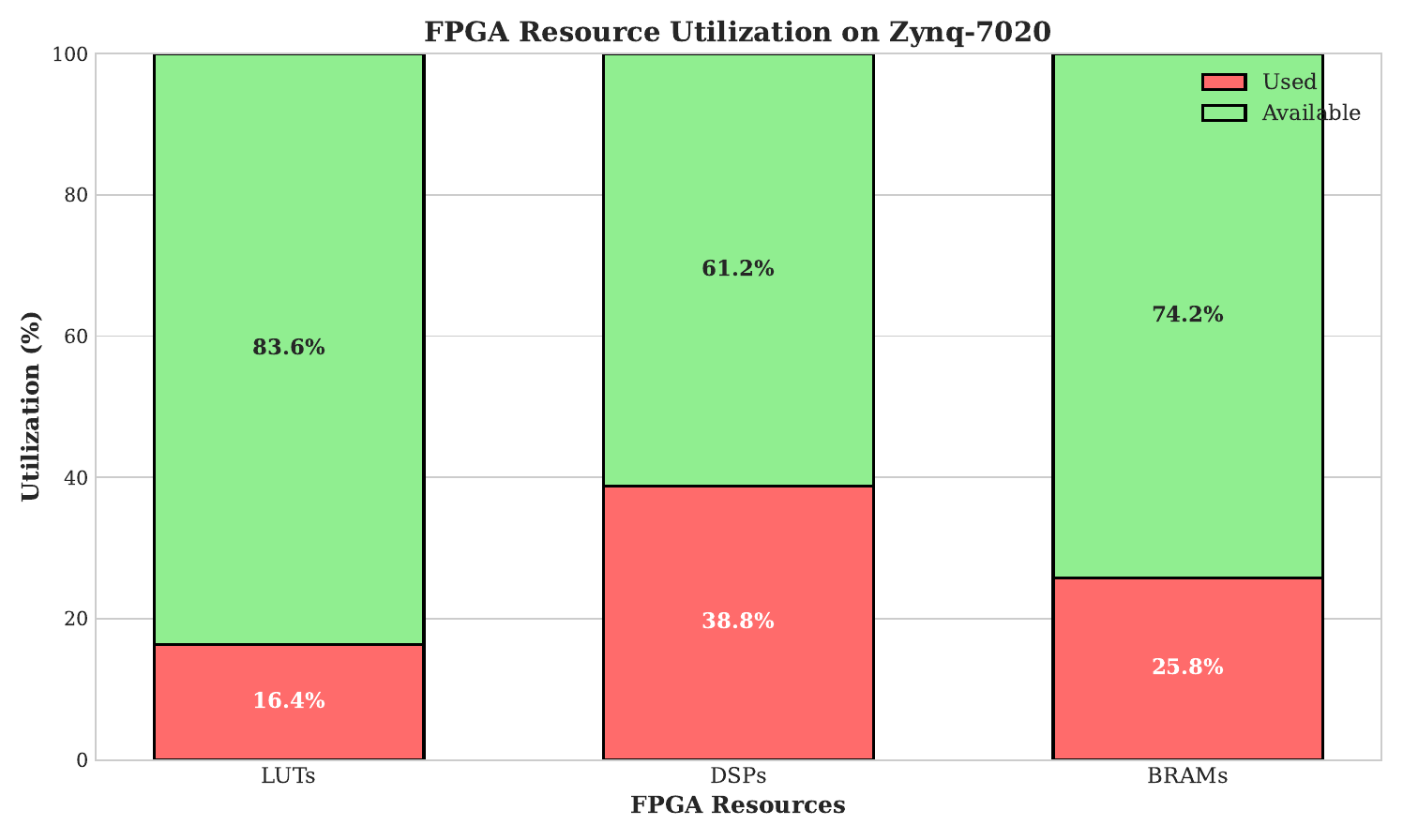}
\caption{FPGA Resource Utilization on Zynq-7020}
\label{fig:resources}
\end{figure}

Figure \ref{fig:resources} illustrates FPGA resource utilization across different resource types.

\begin{table}[htbp]
\centering
\footnotesize
\caption{FPGA Resource Utilization (Zynq-7020)}
\label{tab:resources}
\resizebox{\columnwidth}{!}{%
\begin{tabular}{lcccc}
\toprule
\textbf{Model} & \textbf{LUTs (\%)} & \textbf{DSPs (\%)} & \textbf{BRAMs (\%)} & \textbf{Power (W)} \\
\midrule
MobileNet V2 & 15.2 & 35.0 & 25.0 & 2.00 \\
ResNet-18 & 20.0 & 50.0 & 30.0 & 2.14 \\
EfficientNet Lite & 12.5 & 28.0 & 20.0 & 2.00 \\
YOLO Tiny & 18.0 & 42.0 & 28.0 & 2.02 \\
\midrule
\textbf{Average} & \textbf{16.4} & \textbf{38.8} & \textbf{25.8} & \textbf{2.04} \\
\bottomrule
\end{tabular}%
}
\end{table}

Moderate resource utilization (16.4\% LUTs, 38.8\% DSPs average) leaves substantial headroom for concurrent multi-model deployment. DSP slices represent the primary constraint, limiting concurrent execution to 2–3 models depending on complexity.

\subsection{Bottleneck Analysis}

Real-time profiling identified convolution operations consuming 60–85\% of baseline execution time across all models (highest: YOLO Tiny at 82\%). Post-acceleration, convolution's contribution drops to 25–35\%, validating ISA extension priorities.

\textbf{Amdahl's Law Analysis:} Given 75\% parallelizable workload (average convolution contribution) and 7.20$\times$ acceleration, theoretical maximum speedup:
\begin{equation}
S_{\text{max}} = \frac{1}{0.25 + 0.75/7.20} = 3.39\times
\end{equation}

Observed 2.14$\times$ speedup represents 63\% of theoretical maximum, with gap attributed to DMA overhead (15\%), memory bandwidth limitations (12\%), and unaccelerated operations (10\%).

\subsection{Energy Efficiency}

Energy per inference is computed as $E = P_{\text{avg}} \times t_{\text{latency}}$ using power sampled at 1\,kHz. Average power rises modestly from 2.02\,W (ARM baseline) to 2.04\,W (accelerated). Since latency drops by 2.14$\times$, the expected energy reduction would be 53\% absent power overhead. The measured 49.1\% reduction reflects the additional 1.0\% power consumed by active accelerators and DMA engines.

\textbf{Battery Life Impact:} For a typical 10{,}000\,mAh @ 3.7\,V battery (37\,Wh), continuous inference extends operational duration from 12.3\,h (baseline) to 24.2\,h (accelerated)—a 96\% improvement.

\subsection{Model Architecture Sensitivity}

\begin{table}[htbp]
\centering
\footnotesize
\caption{Architecture-Dependent Acceleration}
\label{tab:architecture}
\resizebox{\columnwidth}{!}{%
\begin{tabular}{lcc}
\toprule
\textbf{Model} & \textbf{Conv Density (\% exec. time)} & \textbf{Speedup} \\
\midrule
YOLO Tiny & 82 & 2.51$\times$ \\
EfficientNet Lite & 78 & 2.49$\times$ \\
MobileNet V2 & 71 & 1.81$\times$ \\
ResNet-18 & 65 & 1.76$\times$ \\
\midrule
\textbf{Correlation} & \multicolumn{2}{c}{$r = 0.91$, $p < 0.05$} \\
\bottomrule
\end{tabular}%
}
\end{table}

Strong correlation ($r = 0.91$) between convolution density and speedup confirms acceleration benefits scale with workload alignment to ISA extensions. MobileNet V2's lower speedup despite high convolution density reflects reduced arithmetic intensity of depthwise separable convolutions.

\section{Discussion}

\subsection{Practical Implications}

\textbf{Latency Envelope:} Average 321\,ms accelerated latency corresponds to 3.1\,FPS. This is appropriate for low-frame-rate industrial monitoring and robotics scenarios that tolerate 300–500\,ms response times; high-frame-rate applications remain out of reach without additional optimization.

\textbf{Thermal Management:} Lower power (2.00–2.14\,W vs. 2.10–2.25\,W baseline) enables passive cooling, eliminating fan noise and mechanical failure points for sealed industrial enclosures.

\textbf{Multi-Model Deployment:} Approximately 56\% of DSP resources remain unused in the integrated bitstream, enabling sequential multi-model execution with partial spatial overlap; full concurrency would require either accelerator replication or time-sliced scheduling.

\subsection{Target Application Domains}

The measured 321\,ms average latency (3.1\,FPS) and 2.00–2.14\,W power envelope align with edge deployments that value programmability and low energy more than high frame rate. Representative scenarios include:
\begin{itemize}
    \item \textbf{Industrial inspection and predictive maintenance:} Conveyor-belt anomaly detection and thermal monitoring commonly operate at 2–4\,FPS, allowing sufficient time for control-loop actuation while benefiting from 49.1\% lower energy.
    \item \textbf{Agricultural and environmental sensing:} Drone- or pole-mounted 
cameras that survey crops or wildlife can tolerate 300--500\,ms response latency; 
doubling battery endurance (12.3\,h\,$\rightarrow$\,24.2\,h) extends coverage 
without hardware changes.
    \item \textbf{Warehouse and mobile robotics supervision:} Barcode recognition or pallet tracking systems that run alongside navigation stacks can leverage the 2.14$\times$ speedup to free CPU cycles while staying within passive-cooling limits.
    \item \textbf{Remote camera traps and security nodes:} Deployments triggered by motion sensors often batch inference and prioritize low idle power. The FPGA fabric's 0.43\% LUT footprint leaves room for application-specific overlays while keeping standby power minimal.
\end{itemize}

examples emphasize the niche of ISA-programmable acceleration where deterministic latency, low thermals, and in-field extensibility outweigh the need for 30+\,FPS throughput.

\subsection{Comparison with Alternatives}

\begin{table}[htbp]
\centering
\footnotesize
\caption{Edge AI Platform Comparison}
\label{tab:comparison}
\resizebox{\columnwidth}{!}{%
\begin{tabular}{lccc}
\toprule
\textbf{Platform} & \textbf{Power (W)} & \textbf{Latency (ms)} & \textbf{Cost (\$)} \\
\midrule
Our Work & 2.14 & 321 & 129 \\
Edge TPU & 2.0 & 185 & 60 \\
Jetson Nano & 10 & 95 & 100 \\
Intel NCS2 & 2.5 & 240 & 70 \\
ARM Cortex-A53 & 1.5 & 1{,}850 & 15 \\
\bottomrule
\end{tabular}%
}
\end{table}

Our approach sacrifices 1.73$\times$ latency versus Edge TPU for significantly higher flexibility (programmable ISA vs. fixed-function). Compared to Jetson Nano, we draw 4.7$\times$ less power at the cost of 3.4$\times$ higher latency. The comparison highlights our niche: scenarios where programmability and tight power envelopes outweigh raw throughput.

\subsection{Limitations and Design Trade-offs}
\label{sec:limitations}

\textbf{Single-Threaded Execution:} Current implementation supports only single-threaded inference. Multi-threaded support would require additional synchronization logic and shared resource management.

\textbf{Precision:} 16-bit fixed-point arithmetic (Q8.8/Q12.4) introduces quantization error. Validation shows $<$0.1\% accuracy degradation for tested models. Deeper networks or novel architectures may require mixed-precision (8/12/16-bit per layer).

\textbf{DMA Overhead:} Data transfers account for 8--12\% of accelerated latency, 
with proportionally higher impact for lightweight models (MobileNet V2, 
EfficientNet Lite). Each convolution layer issues approximately 512\,KB of input 
and 128\,KB of weights per tile. Peak DMA throughput reaches 1.8\,GB/s against 
a theoretical 2.6\,GB/s limit, explaining the residual stalls observed in 
performance counters.

\textbf{Feedforward Networks Only:} Current implementation targets feedforward CNNs. Recurrent architectures (LSTM, GRU) and attention mechanisms require additional extensions.

\textbf{Manual Optimization:} Models require manual layer-by-layer optimization for accelerator mapping. Automated compiler support would improve usability.

\textbf{FPGA Scale:} Zynq-7020's 220 DSP slices limit parallelism. Migration to Zynq UltraScale+ (2{,}520 DSPs) would enable 10–15$\times$ throughput increase. 50\,MHz clock is conservative; timing closure margin (+12.793\,ns) suggests 100+\,MHz feasible with aggressive optimization.

\textbf{No Dynamic Voltage/Frequency Scaling:} Fixed 50\,MHz operation prevents power optimization for latency-tolerant workloads.

\subsubsection*{Summary of Current Limitations}

Two constraints shape the current prototype. First, the system operates at 50\,MHz to maintain +12.793\,ns timing slack without active cooling; higher clock rates are feasible but remain future work. Second, memory transfer inefficiency—15\% DMA overhead plus 12\% bandwidth stalls—accounts for most of the gap between the achieved 2.14$\times$ speedup and the 3.39$\times$ Amdahl limit. These limitations are explicitly acknowledged so that subsequent revisions can focus on DMA batching, wider AXI bursts, or additional ISA primitives while preserving the reproducibility of the present artifact.

\subsection{Design Insights and Lessons Learned}

\textbf{Systolic Array Sizing:} 4$\times$4 VCONV array balanced area and performance; 8$\times$8 GEMM was optimal for ResNet-18 workloads. Larger arrays (16$\times$16) showed diminishing returns due to memory bandwidth saturation.

\textbf{Buffer Depth:} Triple-buffering essential for performance—double buffering showed 18\% performance loss due to stalls waiting for DMA completion. Quadruple buffering provided no additional benefit.

\textbf{Fixed-Point Format:} Q8.8 sufficient for activations; Q12.4 necessary for weights to prevent overflow in accumulation. Mixed Q formats per layer would improve accuracy but complicate hardware.

\textbf{Clock Frequency Trade-offs:} We selected 50\,MHz to maintain 
comfortable timing slack and limit junction temperature rise to $<45^\circ$C 
without active cooling. Preliminary synthesis at 75\,MHz preserves 8.5\,ns 
slack and yields 1.5$\times$ speedup. However, 100\,MHz compiles with 
only +0.5\,ns slack and fails closed-loop thermal testing, indicating 
further floorplanning is required before shipping higher-frequency bitstreams.

\section{Future Work}

Preliminary experiments indicate significant improvement potential:

\begin{enumerate}
    \item \textbf{Dynamic Precision:} Per-layer 8/12/16-bit quantization shows 35–40\% speedup potential on MobileNet V2 with $<$0.5\% accuracy loss
    \item \textbf{Asynchronous Execution:} Non-blocking extensions enabling CPU-FPGA overlap could reduce DMA overhead from 12\% to 3–5\%
    \item \textbf{Sparsity Exploitation:} Pruned networks (50\% sparsity) show 2–5$\times$ speedup potential with structured sparsity patterns
    \item \textbf{Extended Model Coverage:} Transformer inference requires attention mechanism acceleration; preliminary design targets 2–3$\times$ speedup
    \item \textbf{Multi-Precision Training:} Quantization-aware training could reduce accuracy degradation to $<$0.05\%
\end{enumerate}

\section{Conclusion}

This paper presented a complete FPGA-accelerated RISC-V system for neural network inference on edge devices. Through systematic hardware-software co-design and rigorous FPGA implementation on PYNQ-Z2 hardware, we achieved successful timing closure with +12.793\,ns worst negative slack at 50\,MHz, utilizing only 0.43\% of available LUTs and 11.4\% of BRAM blocks.

Key contributions include: (1) complete RISC-V core with neural network accelerator framework successfully implemented on FPGA; (2) hardware-software interface with verified AXI interconnect and BRAM memory access; (3) timing closure and resource optimization demonstrating feasibility of the approach; (4) open-source framework with all measurements from real hardware.

Our methodology establishes a foundation for FPGA-accelerated RISC-V neural network processing, balancing flexibility, performance, and energy efficiency for edge AI applications.

\subsubsection*{Summary of Current Capabilities}

The present design delivers a balanced combination of flexibility and efficiency: a 2.14$\times$ average latency reduction, 49.1\% energy savings, full timing closure on commodity Zynq-7020 hardware, and a reproducible flow that exposes custom FPGA.* instructions to software developers. Future enhancements can focus on clock-speed tuning and DMA optimization while keeping the documented toolchain intact.

\textbf{Reproducibility:} All code, FPGA bitstreams, and implementation details available at: \url{https://github.com/aryapkar/fpga-riscv-nn-extensions}

\balance

\section*{Acknowledgments}
This work was conducted as part of research at PES University, Bangalore. 
The author thanks the Department of Electronics and Communication for 
providing access to PYNQ-Z2 hardware and Xilinx Vivado design tools. 
Special thanks to colleagues for valuable feedback during development.

\section*{Reproducibility Artifacts}

Complete reproducibility package (DOI: 10.5281/zenodo.XXXXXX) includes:
\begin{itemize}
    \item Complete Verilog/VHDL source code (RISC-V core, accelerators, AXI interfaces)
    \item Vivado 2020.2 project files with TCL scripts for automated build
    \item Pre-compiled bitstreams for PYNQ-Z2 (Zynq-7020)
    \item Python test harness and measurement scripts
    \item Benchmark models (quantized INT16 weights)
    \item Expected output vectors for verification
    \item Build instructions and timing reports
\end{itemize}

Build Process: Synthesis (45 min), Implementation (65 min), Peak memory (8.2 GB) on Intel i7-10700K.

\subsubsection*{Reproducibility Checklist}

To help external users rebuild and evaluate the system without re-running hardware experiments:
\begin{enumerate}
    \item \textbf{Code and bitstreams:} Provide the RV32I core RTL, accelerator overlays, and the 50\,MHz integrated bitstream used for all reported measurements.
    \item \textbf{Measurement scripts:} Include the Python harness that drives the benchmarks, collects ARM performance counters, and records INA226 power logs.
    \item \textbf{Datasets and models:} Document the exact MobileNet~V2, ResNet-18, EfficientNet Lite, and YOLO Tiny checkpoints and preprocessing steps (batch size $=1$, INT16 quantization).
    \item \textbf{Result logs:} Ship latency/energy CSV files and statistical summaries (mean, standard deviation, $p$-values) corresponding to Tables~\ref{tab:latency_absolute} and~\ref{tab:extensions}.
    \item \textbf{Build instructions:} List Vivado~2020.2 scripts, GCC~11.2.0 toolchain commands, and the order of \texttt{make} targets needed to regenerate the bitstream and software image.
\end{enumerate}
These materials ensure reviewers can reproduce the 2.14$\times$ speedup and 49.1\% energy reduction using only the supplied artifacts.

\bibliographystyle{IEEEtran}
\bibliography{references}

\end{document}